\documentclass{aa}
\usepackage{graphicx}
\usepackage{amssymb}
\usepackage{tabularx}
\usepackage[figuresright]{rotating}
\def\aips{${\cal A}{\cal I}{\cal P}{\cal S}$}

\begin{document}
   \title{The Twin--Jet System in NGC\,1052: \\
VLBI-Scrutiny of the Obscuring Torus}


   \author{M.\,Kadler
          \inst{1}
          \and
          E.\,Ros
\inst{1}
          \and
	  A.\,P.\,Lobanov
\inst{1}
          \and
	  H.\,Falcke\inst{1}\fnmsep\thanks{Present Address: ASTRON, P.O. Box 2, 7990 AA Dwingeloo, The Netherlands}
	  \and
	  J.\,A.\,Zensus
\inst{1} 
          }

   \offprints{{\tt mkadler@mpifr-bonn.mpg.de}}

   \institute{Max-Planck-Institut f\"ur Radioastronomie,
              Auf dem H\"ugel 69, 53121 Bonn, Germany\\
             }

   \date{Received 8 April 2004; accepted 8 July 2004}

   \abstract{NGC\,1052 offers the possibility to study the obscuring torus
around a supermassive black hole,
 predicted by the standard model of active galactic nuclei,
over a wide range of wavelengths
from the radio to the X-ray regime. We present
a detailed VLBI study of the parsec-scale structure of the ``twin-jet'' 
system in NGC\,1052 in both total and polarized intensity and at multiple
frequencies. We report the detection of linearly polarized emission from the
base of the eastern jet at 5\,GHz. While the radio spectrum in this region
might be still consistent with synchrotron self absorption, the highly
inverted spectrum of the western jet base represents a clear sign of 
pronounced free-free absorption in a circumnuclear torus. We observe
an abrupt change of the brightness temperature gradient at a distance of 
$\sim 0.2$\,pc to $0.3$\,pc from the central engine. This might provide
an observational signature of the edge of the central torus, where the transition 
from an external pressure-dominated jet regime to a more or less freely expanding
jet takes place.
We determine the absorbing column density towards the western jet core
to be $\sim 2.2 \times 10^{22}$\,cm$^{-2}$ in good agreement with the
values derived from various X-ray observations. This suggests
that the nuclear X-ray emission and the jet emission imaged by VLBI originate
on the same scales.
   }

\titlerunning{The Twin--Jet System in NGC\,1052}
   \maketitle
%

\section{Introduction}
Radio observations of the low-luminosity
active galactic nucleus (AGN) NGC\,1052
with Very Long Baseline Interferometry (VLBI) at multiple frequencies
have revealed the presence of a dense circumnuclear absorber, which
obscures the very center of this elliptical galaxy 
(Kellermann et al. \cite{Kel99}; Kameno et al. \cite{Kam01}; Vermeulen et al.
\cite{Ver03}). 
Indeed, the standard model of AGNs predicts
the existence of an obscuring torus, whose inner surface is expected to be
photo-ionized by illumination from the accretion disk. Colder, neutral 
material forms the outer boundary of the obscuring torus.
NGC\,1052 provides the possibility to study the physical properties of this
obscuring torus complementary in various wavelength regimes and with a variety
of observational methods. Particularly, the combination of VLBI and X-ray 
spectroscopic studies is capable of addressing the same basic questions with
complementary methods.  
Various
X-ray observations of NGC\,1052 imply a model-dependent
column density of $10^{22}$\,cm$^{-2}$ to $10^{23}$\,cm$^{-2}$ towards the
unresolved nuclear
X-ray core (Guainazzi et al. \cite{Gua99}; Weaver et al. \cite{Wea99};
Kadler et al. \cite{Kad04})
but the angular resolution of X-ray telescopes
is not sufficient to measure the accurate position and extent of the
absorber.
Relativistically broad iron line emission at 6.4\,keV is seen in a high-quality
X-ray spectrum obtained with the {\it XMM-Newton} telescope (Kadler et al. in prep.).
Thus, being the first radio-loud AGN with a strong compact radio core that exhibits
{ strong
relativistically broadened iron line emission from the inner accretion disk}, 
NGC\,1052 provides a unique possibility to study
the inter-relation between AGN mass accretion and jet-formation. 
For future combined VLBI structural and X-ray spectroscopic
monitoring observations, it is essential to study in detail the influence of the
obscuring torus on the parsec-scale jet structure at radio wavelengths.

NGC\,1052 is a moderately strong, variable source 
in the radio regime, with a luminosity
(integrated between 1\,GHz and 100\,GHz) of $\sim4.4 \cdot 10^{40}$\,erg\,s$^{-1}$
(Wrobel \cite{Wro84}) 
and an unusually bright and compact radio core.
Together with
the proximity of the source of only 22.6\,Mpc\footnote{We assume a Hubble
constant of $H_0=65$\,km\,s$^{-1}$\,Mpc$^{-1}$ and use the measured redshift
of $z=0.0049$ by Knapp et al. (\cite{Kna78}). This results to a
linear scale of 0.11\,pc\,mas$^{-1}$.}  this makes NGC\,1052 a
premier object for VLBI studies, aiming at the ultimate goal of revealing
the physical properties of obscuring tori in AGNs. 
The parsec-scale structure of NGC\,1052 
shows a twin jet with an emission gap between the brighter (approaching)
eastern jet and the western (receding) jet  
and free-free absorption towards the western jet 
(Kellermann et al. \cite{Kel99}).
Kameno et al. (\cite{Kam01}) suggested the presence of a geometrically
thick plasma torus and
a geometry of the jet-torus system in which 0.1\,pc of the eastern jet and 0.7\,pc
of the western jet are obscured.
{ The multi-frequency VLBI
structure of NGC\,1052 has been further studied 
by Kameno et al. (\cite{Kam03}) and Vermeulen et al. (\cite{Ver03}).} 
The kinematics of both jets at 2\,cm have been
investigated
by Vermeulen et al. (\cite{Ver03}) who report outward
motions on both sides of the gap with similar velocities around
0.6 to 0.7\,mas yr$^{-1}$ corresponding to $\sim 0.25\,c$. 

Besides the ionized
(free-free absorbing) gas component there are multiple pieces of evidence for
atomic and molecular gas in the central region of NGC\,1052.
H$_2$O maser emission occurs towards the base of
the western jet (Claussen et al. \cite{Cla98}) within the same region 
that is heavily affected
by free-free absorption. 
Atomic hydrogen is known to exist in NGC\,1052 on various scales.
Van Gorkom et al. (\cite{VanG86}) imaged the distribution of the H{\sc i} gas with a
resolution of $1^{\prime\prime}$ using the VLA. They report a structure three
times the size of the optical galaxy.
Recent VLBI
observations resolved H{\sc i} absorption features towards the
nuclear jet (Vermeulen et al. \cite{Ver03}). 
Finally, an
OH absorption line was also detected by Omar et al. (\cite{Oma02}) { and
Vermeulen et al. (\cite{Ver03})}
but the distribution of
the OH gas on parsec-scales has not been investigated so far.

In this work we analyse the multi-frequency structure of NGC\,1052
on parsec-scales between 5\,GHz and 43\,GHz in both total and linearly
polarized intensity. In Sect.~\ref{obs} we describe briefly the observations
with the Very Long Baseline Array (VLBA)\footnote{Napier (\cite{Nap94});
the VLBA is operated by the National Radio Astronomy Observatory (NRAO),
a facility of the National Science Foundation operated under cooperative
agreement by Associated Universities, Inc.} 
and the data reduction.
Section~\ref{modeling} and Sect.~\ref{aligning} describe the modeling of the source
structure with Gaussian components and the process of aligning the images at
the four frequencies.
The VLBA images of NGC\,1052 themselves in total and polarized intensity
are presented in 
Sect.~\ref{maps}. 
We analyse the frequency dependence of the
observed core position in both jets in Sect.~\ref{coreshift} and
Sect.~\ref{spectral} discusses the spectral analysis. 
In Sect.~\ref{t_b} we present the brightness temperature distribution
along both jets of NGC\,1052 and
summarize our results in Sect.~\ref{sum}. 

\section{Observations and data reduction}
\label{obs}
NGC\,1052 was observed on December 28th, 1998 with the
VLBA
at four frequencies (5\,GHz, 8.4\,GHz, 22\,GHz, and 43\,GHz) in dual polarization mode.
The data were recorded with
a bit rate of 128\,Mbps at 2-bit sampling providing a bandwidth of 16\,MHz
per polarization hand (divided in two blocks of 16 0.5\,MHz channels each). The total 
integration time on NGC\,1052 was about one hour at 5\,GHz, 8.4\,GHz, and
22\,GHz each, and about six hours at 43\,GHz, to compensate the lower
array sensitivity and the lower source flux density.
3C\,345 and {\rm 4C\,28.07} were used as calibrators during the observation.
The correlation of the data was done at the Array Operations Center of the VLBA in
Socorro, NM, USA, with an averaging time of two seconds.
All antennas of the array yielded good data, except 
the Owens Valley antenna, which did not record data at 8.4\,GHz and 22\,GHz.
The data at 22\,GHz and 43\,GHz suffered from some snow in
the
Brewster dish and rain at St. Croix.

The data calibration and imaging were performed applying standard methods using
the programs \aips\ 
and {\sc difmap} (Shepherd et al. \cite{She97}). 
The {\it a-priori} data calibration
and fringe fitting were performed in \aips\ using the nominal gain curves
measured for each antenna. 
Instrumental phase offsets and gradients were corrected using the 
phase-cal signals injected into the data stream during the data recording
process. 
The data were averaged over frequency and
exported from \aips . Then, the data were read into {\sc difmap}, 
edited, phase- and
amplitude self-calibrated, and imaged { by making use of the {\sc clean} 
algorithm. From a careful comparison of the uncalibrated and self-calibrated data,
the absolute flux calibration can be (conservatively) estimated to be accurate 
on a level of $\lesssim 10$ percent.}

The data were corrected for the instrumental polarization of the
VLBA using the method
described by Lepp\"anen et al. (\cite{Lep95}).
The absolute values
of the electric vector position angles (EVPAs)
at all four frequencies were calibrated using { the source 3C\,345},
for which a large data base
exists in the literature from which the EVPA in the core and jet regions of
this source can be obtained (e.g., Ros et al. \cite{Ros00}).

\section{Modeling the source structure}
\label{modeling}
In VLBI imaging the absolute positional information is lost
in the phase-calibration process. In the case of simultaneous multi-frequency
observations this means that {\it a--priori} it is not clear how the images at the
different frequencies have to be aligned. The ideal method for overcoming this
is to carry out phase-referencing observations, using a compact nearby object
and using it to calculate the position of the target source relative to it
(e.g., Ros \cite{Ros04}).
Our VLBA observations were not phase-referenced, so another way was used
to register the four maps.
We model fitted the visibility data with two-dimensional Gaussian
functions. Those functions, called components, were chosen to be circular 
to reduce the { number of free model parameters} and to facilitate the
comparison of the models at the different frequencies.
The model fitting was performed in {\sc difmap}
using the least-squares method.
The errors were determined with the program
{\sc erfit}, a program from the Caltech VLBI data analysis package 
that
calculates the statistical confidence intervals of the fitted model
parameters by varying each parameter.

The fits were initially performed independently at the different frequencies 
to avoid biasing one by another. Once good fits for all four frequencies were
obtained, a cross comparison of the resulting maps was made and the fits were
modified to get a set of model fits as consistent as possible.
Criteria for consistency were:

\begin{itemize}
\item regions that show emission at adjacent frequencies should be
represented by the same number of components,
\item extended components in the outer parts of the jets
should become weaker at higher frequencies as they most likely represent optically thin
synchrotron emitting regions,
\item the inner edges of both jets 
are expected to shift inwards towards higher frequencies due to opacity effects, i.e.,
the corresponding model components might have no low frequency counterpart,
\item optically-thin features should not show positional changes with frequency.
\end{itemize}

Table \ref{tab:fits} gives the parameters of the final models
for the four frequencies.
The most distant component in the eastern jet was labeled
as A\,1, the adjacent inner one as A\,2, and so on. The western jet was divided
into three parts following the convention introduced above. The
model fit component in the innermost part of the western jet were labeled as
B\,2b, B\,2a, B\,1 from east to west.
Further out, the components C3b, C3a, C2,
C1, and D follow.

\begin{table}
\begin{flushleft}
\caption{Model fit parameters}
\label{tab:fits}
\begin{center}
\[
\centering
\centerline{
\resizebox{0.79\columnwidth}{!}{%
\begin{tabular}{@{}c@{~}c@{~}c@{~}c@{~}c@{}}
\hline
\hline
\noalign{\smallskip}
    & Flux    \\
 Id & density & Radius$^{\rm a}$ & P.A. & FWHM \\ 
    & [mJy]   &  [mas]           & [$^\circ$]& [mas] \\ 
\noalign{\smallskip}
\hline
\noalign{\smallskip}
\multicolumn{5}{@{}c@{}}{
\it 5\,GHz, $\chi_{\rm red}=1.12$ \hrulefill 
} \\
\noalign{\smallskip}
A1  & 80$\pm$2    & 14.56$\pm$0.03 & 66.5$\pm$0.1 & 2.14$\pm$0.03  \\
A2  & 54$\pm$2    & 12.01$\pm$0.04 & 63.6$\pm$0.1 & 1.87$\pm$0.08  \\
A3  & 60$\pm$2    & 9.61$\pm$0.05 &  65.3$\pm$0.3 & 1.63$\pm$0.04  \\
A4  & 150$\pm$7   & 7.46$\pm$0.05 &  71.7$\pm$0.2 & 0.99$\pm$0.02  \\
A5  & 338$\pm$37  & 6.0$\pm$0.2 &    72.8$\pm$0.1 & 0.93$\pm$0.25  \\
A6  & 447$\pm$48  & 5.1$\pm$0.1 &    72.8$\pm$0.2 & 0.83$\pm$0.19  \\
A7/8& 450$\pm$28 &  4.0$\pm$0.1 &    71.0$\pm$0.2 & 0.69$\pm$0.08  \\
A9  & 397$\pm$30  & 3.1$\pm$0.2 &    71.1$\pm$0.3 &  0.20$\pm$0.07  \\
A10 & 140$\pm$37  & 2.3$\pm$0.2 &    70.9$\pm$0.4 &  0.46$\pm$0.05  \\ 
C3  & 137$\pm$3   & 3.84$\pm$0.03 &  $-114.2$$\pm$0.1 & 1.03$\pm$0.06  \\
C2  & 59$\pm$3    & 4.75$\pm$0.06 &  $-115.74$$\pm$0.3 & 0.83$\pm$0.09  \\
C1  & 24$\pm$1    & 7.19$\pm$0.03 &  $-119.6$$\pm$0.4 & 1.85$\pm$0.10  \\
D   & 67$\pm$1    & 11.81$\pm$0.01 & $-114.7$$\pm$0.1 & 2.24$\pm$0.05  \\
\multicolumn{5}{@{}c@{}}{
\it 8.4\,GHz, $\chi_{\rm red}=1.76$ \hrulefill 
} \\
\noalign{\smallskip}
A1  &   38$\pm$1 & 14.70$\pm$0.03 & 66.5$\pm$0.1 &  2.00$\pm$0.03
\\
A2  &   37$\pm$1 & 12.04$\pm$0.04 & 64.0$\pm$0.1 &  2.34$\pm$0.08
\\
A3  &   36$\pm$1 & 9.30$\pm$0.05 & 65.9$\pm$0.3 &  1.62$\pm$0.04 \\
A4  &   83$\pm$4 & 7.34$\pm$0.05 & 71.9$\pm$0.2 &  0.98$\pm$0.02 \\
A5  &   72$\pm$8 & 6.4$\pm$0.2 & 72.7$\pm$0.1 &  0.59$\pm$0.25 \\
A6  &   360$\pm$26 & 5.5$\pm$0.1 & 72.8$\pm$0.2 & 0.81$\pm$0.12 \\A7  &   396$\pm$22 & 4.4$\pm$0.1 & 72.3$\pm$0.2 & 0.74$\pm$0.06 \\A8  &   146$\pm$8 &  3.8$\pm$0.1 & 69.7$\pm$0.2 & 0.28$\pm$0.06 \\A9  &   360$\pm$21 & 3.2$\pm$0.2 & 71.3$\pm$0.3 & 0.34$\pm$0.05 \\A10 &   316$\pm$47 & 2.7$\pm$0.2 & 69.2$\pm$0.4 & 0.31$\pm$0.03 \\A11 &   137$\pm$25 & 1.9$\pm$0.2 & 74.3$\pm$0.4 & 0.34$\pm$0.07 \\B2a/b & 56$\pm$4 & 0.60$\pm$0.03 & $-119.8$$\pm$0.1 &  0.36$\pm$0.02 \\
B1  &   29$\pm$2 & 1.26$\pm$0.03 & $-119.9$$\pm$0.1 &  0.31$\pm$0.02 \\
C3  &   20$\pm$1 & 3.75$\pm$0.03 & $-114.2$$\pm$0.1 &  1.06$\pm$0.08 \\
C2  &   77$\pm$4 & 4.86$\pm$0.06 & $-114.9$$\pm$0.3 &  0.90$\pm$0.09 \\
C1  &   21$\pm$1 & 7.03$\pm$0.03 & $-119.1$$\pm$0.4 &  1.68$\pm$0.10 \\
D   &   48$\pm$1 & 11.95$\pm$0.01 & $-114.7$$\pm$0.1 &  2.21$\pm$0.05 \\
\multicolumn{5}{@{}c@{}}{
\it 22\,GHz, $\chi_{\rm red}=0.71$ \hrulefill 
} \\
\noalign{\smallskip}
A6  &  138$\pm$5 & 5.48$\pm$0.03 & 72.4$\pm$0.2 & 0.98$\pm$0.04
\\
A7  &  140$\pm$7 & 4.39$\pm$0.01 & 72.2$\pm$0.2 & 0.65$\pm$0.03\\
A8  &  99$\pm$5  & 3.72$\pm$0.01 & 68.1$\pm$0.2 & 0.47$\pm$0.03 \\A9  &  139$\pm$6 & 3.09$\pm$0.01 & 72.2$\pm$0.2 & 0.32$\pm$0.02 \\A10 &  144$\pm$5 & 2.70$\pm$0.01 & 68.5$\pm$0.2 &  0.27$\pm$0.01 \\
A11 &  44$\pm$8  & 1.98$\pm$0.04 & 73.0$\pm$0.8 & 0.29$\pm$0.07 \\A12 &  86$\pm$15 & 1.53$\pm$0.04 & 72.1$\pm$0.4 & 0.26$\pm$0.06 \\A13 &  54$\pm$14 & 1.20$\pm$0.04 & 71.1$\pm$0.9 & 0.22$\pm$0.06 \\B2b &  341$\pm$14& 0.52$\pm$0.02 & $-123$$\pm$1 & 0.21$\pm$0.01 \\
B2a &  151$\pm$14& 0.72$\pm$0.05 & $-115$$\pm$2 & 0.29$\pm$0.02 \\
B1  &  31$\pm$2  & 1.38$\pm$0.02 & $-132$$\pm$1 & 0.3$\pm$1.7 \\
C3a & 50$\pm$4   & 3.34$\pm$0.02 & $-116.4$$\pm$0.4 & 0.58$\pm$0.03 \\
C3b & 86$\pm$6   & 4.18$\pm$0.07 & $-116$$\pm$1 & 1.95$\pm$0.16 \\
\multicolumn{5}{@{}c@{}}{
\it 43\,GHz, $\chi_{\rm red}=0.83$ \hrulefill 
} \\
\noalign{\smallskip}
A6  & 40$\pm$2 & 5.36$\pm$0.03 &   72.8$\pm$0.2 &  0.92$\pm$0.05 \\
A7  & 44$\pm$2 & 4.39$\pm$0.01 &   72.2$\pm$0.2 & 0.51$\pm$0.03 \\A8  & 39$\pm$2 & 3.76$\pm$0.01 &   68.7$\pm$0.1 & 0.46$\pm$0.03 \\A9  & 37$\pm$1 & 3.177$\pm$0.003 & 71.8$\pm$0.1 & 0.22$\pm$0.01  \\
A10 & 70$\pm$1 & 2.772$\pm$0.004 & 68.4$\pm$0.1 & 0.33$\pm$0.01  \\
A11 & 19$\pm$1 & 1.95$\pm$0.01 &   71.6$\pm$0.5 &  0.32$\pm$0.03 \\
A12 & 57$\pm$2 & 1.499$\pm$0.004 & 70.2$\pm$0.2 & 0.22$\pm$0.01 \\A13 & 43$\pm$2 & 1.221$\pm$0.006 & 70.8$\pm$0.2 &  0.12$\pm$0.01 \\
A14 & 54$\pm$1 & 1.017$\pm$0.003 & 71.9$\pm$0.2 & 0.08$\pm$0.05 \\A15 & 19$\pm$1 & 0.47$\pm$0.01 &   61$\pm$2     & 0.21$\pm$0.03 \\B2b & 225$\pm$2 & 0.467$\pm$0.002 &$-119.2$$\pm$0.3 &  0.224$\pm$0.002 \\
B2a & 31$\pm$2 & 0.779$\pm$0.007 & $-120.9$$\pm$0.6 &  0.22$\pm$0.01 \\
C3b & 27$\pm$2 & 3.33$\pm$0.05 &   $-116.5$$\pm$0.8 &   0.95$\pm$0.08 \\
\noalign{\smallskip}
\hline
\end{tabular}
}}
\]
\end{center}
\begin{list}{}{
\setlength{\leftmargin}{0pt}
\setlength{\rightmargin}{0pt}
}
\item[$^{\mathrm{a}}$] The radius is measured from the center between A15 and B2b (see Section 4.6).
\end{list}
\end{flushleft}
\end{table}

\section{Image alignment}
\label{aligning}
   \begin{figure*}[t!]
\vbox{\includegraphics[clip, width=12cm]{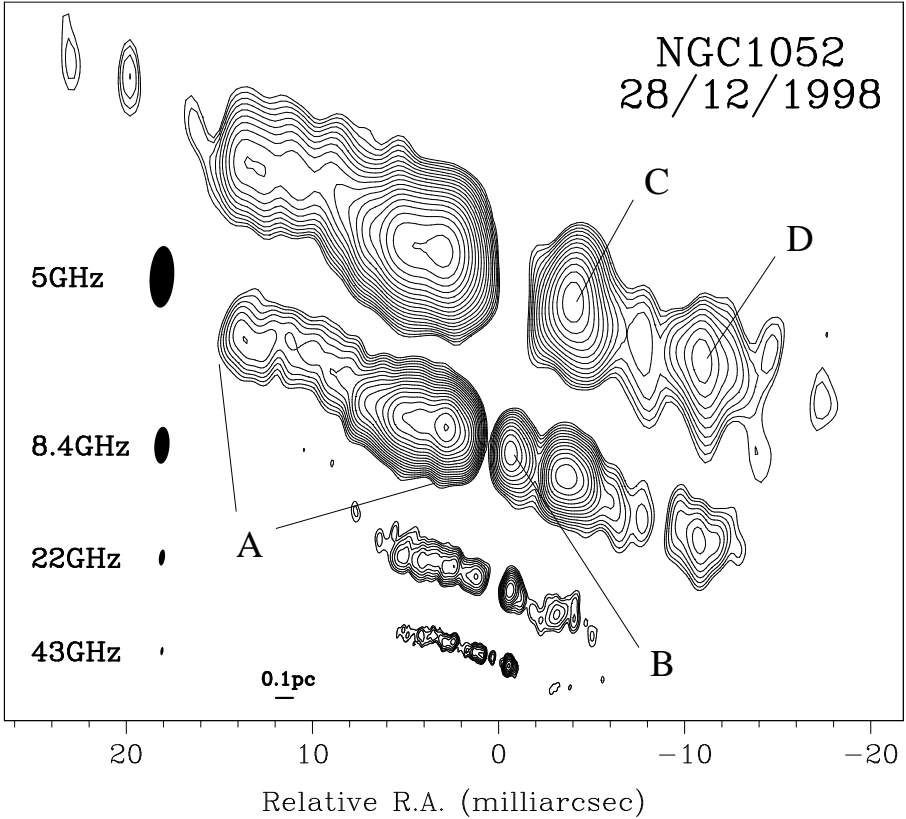}}\vspace{-2cm}
\hfill \parbox[b]{5.5cm}
{\caption{The aligned VLBA images of NGC\,1052 at 5\,GHz, 8.4\,GHz, 22\,GHz, and 43\,GHz.
Contours and beam sizes are
given in Table \ref{tab:4maps}.
              \label{fig:aligned}}}
    \end{figure*}
\begin{figure*}[htb]
\vbox{\includegraphics[clip, width=12cm]{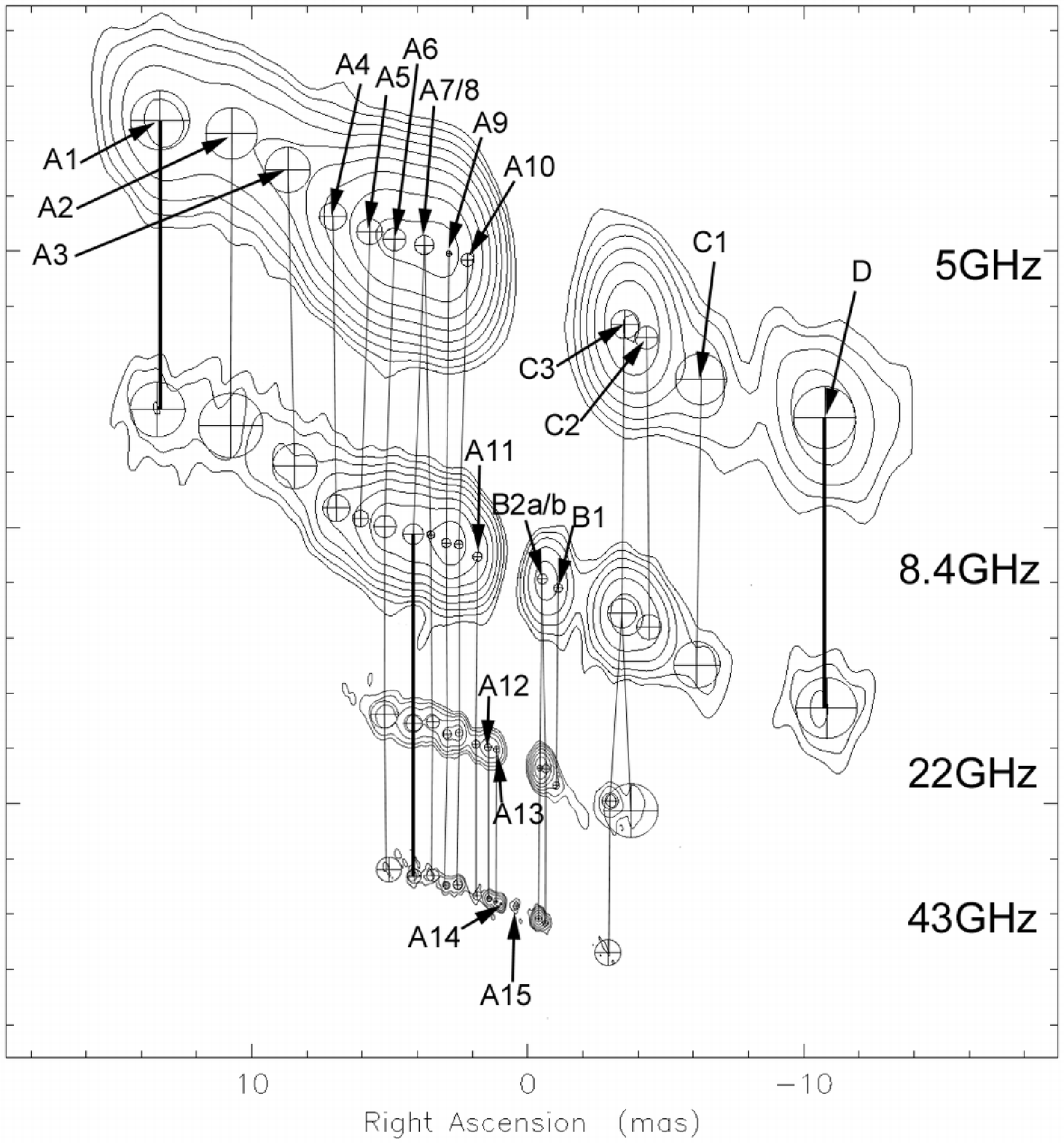}}\vspace{-6cm}
\hfill \parbox[b]{5.5cm}
{\caption{The aligned model-fit maps. The circles represent the
full width at half maximum (FWHM) of the circular
Gaussians. The vertical and oblique lines join the associated components between
different frequencies. The components A\,1 and D between 5\,GHz and 8.4\,GHz and A\,7
between
8.4\,GHz, 22\,GHz and 43\,GHz (joined by a thicker trace) were used for the alignment.
The lowest
contours are 1.9\,mJy/beam for the 5\,GHz model,
2.6\,mJy/beam for the 8.4\,GHz model,
5.0\,mJy/beam for the 22\,GHz model and
3.5\,mJy/beam for the 43\,GHz model.
\label{fig:allmodels}}}
\end{figure*}

The image alignment was performed in two steps.
{\sc i)} The fitted components at the different frequencies
were cross--identified and the relative shifts between the models
were determined, assuming frequency-independent
positions of optically thin features.
{\sc ii)} An origin was determined from which absolute distances could be 
measured.
This method of alignment is not {\it a--priori} definite. It is based
on the assumption that optically-thin components have frequency-independent
positions and that the cross identification of the model components is correct.
{ However, all possible
alternative identifications could be ruled out for consistency reasons (by making
use of the component flux densities at adjacent frequencies).}

We used the two most distant components (A\,1 and D)
to align the 5\,GHz and the 8.4\,GHz models by assuming that the mid-point between both
components was spatially coincident. Because these two outer components are
optically thin and not detectable above 8.4\,GHz,  
the position of component A\,7, 
which is relatively strong at all three frequencies,
was used to align the three high-frequency models relative to the origin determined
from the alignment of the two low-frequency models.
As a natural choice, the most probable
position of the true center of jet activity, namely the
center between the components A\,15 and B\,2b, was used as the origin 
(compare Sect.~\ref{coreshift}). 
The component
positions in Table \ref{tab:fits} are given relative to
this reference point.

\section{The brightness distribution in total and polarized intensity}
\label{maps}
\subsection{Total intensity imaging}
Figure~\ref{fig:aligned} shows the aligned uniformly weighted total intensity images of NGC\,1052
at the four frequencies and Fig.~\ref{fig:allmodels} shows the
images that result from applying Gaussian model-fit components to the measured
visibilities. Table \ref{tab:4maps} gives the image parameters of Fig.~\ref{fig:aligned}. 

The basic source structure is formed of
two oppositely directed jets divided by an emission gap.
The source can be divided into four regions: region A makes up
the whole (more or less continuous) eastward-directed jet emission;
region B is not visible at 5\,GHz, but becomes the brightest
and most compact feature at high frequencies; regions C and D
are bright only at low frequencies and become faint and diffuse
at high frequencies. 
The main source characteristics are summarized in Table \ref{tab:characteristics}.
A sub-division can be performed based on the Gaussian model
fits, found to represent the source structure (see 
Fig.~\ref{fig:allmodels}). Region A is sub-divided into the components
A\,1 to A\,15, region B into the components B\,1, B\,2a, and B\,2b, and region
C into the components C\,1, C\,2, and C\,3.
 
Whilst
the eastern jet
is only slightly curved at distances from the gap larger than
4\,mas
the counterjet exhibits 
strong curvature
most pronounced at 22\,GHz. 
The jets appear nearly symmetric in the 5 GHz and 8.4\,GHz images,
becoming asymmetric at higher frequencies.
While the emission gap is most prominent
at 5\,GHz the images at 8.4\,GHz, 22\,GHz, and 43\,GHz reveal jet components occupying
this gap region, leaving a smaller but still 
prominent emission gap. 
The frequency dependence of the VLBI ``core'' positions, the points were the
eastern and the western jet become optically thin
can clearly be seen in Fig.~\ref{fig:aligned}. This so-called 
``core shift'' will be analysed in detail in Sect.~\ref{coreshift}.

\begin{table}
\caption{Total intensity image parameters}
\label{tab:4maps}
\[
\centering
\resizebox{\columnwidth}{!}{%
\begin{tabular}{@{}c@{~\,}c@{~\,}c@{~\,}c@{~\,}c@{~\,}c@{}}
\hline
\hline
\noalign{\smallskip}
 $\nu$ &  beam  &  $S_{\rm peak}$  &  \multicolumn{1}{@{}c@{}}{$S_{\rm tot}$$^{\rm (a)}$} &  rms &  Lowest Contour \\

{\scriptsize [GHz]} & {\scriptsize[mas $\times$ mas,$^\circ$]} & {\tiny [Jy/beam]}  & {\scriptsize [Jy]} & \multicolumn{1}{@{}c@{}}{\tiny [mJy/beam]} & {\scriptsize [mJy/beam]} \\ 
\noalign{\smallskip}
\hline
\noalign{\smallskip}
5 & 3.30$\times$1.31,\,$-3.74$ & 0.660$^{\rm b}$ & 2.41 & 0.26 & 1.5 \\
8.4 & 1.98$\times$0.81,\,$-3.87$ & 0.538$^{\rm b}$ & 2.39 & 0.25 & 1.5 \\
22 & 0.86$\times$0.32,\,$-7.63$ & 0.339$^{\rm c}$ & 1.51 & 1.20 & 2.5 \\
43 & 0.45$\times$0.16,\,$-7.93$ & 0.126$^{\rm c}$ & 0.67 & 0.67 & 2.5 \\
\noalign{\smallskip}
\hline
\end{tabular}
}
\]
\begin{footnotesize}
\begin{list}{}{
\setlength{\leftmargin}{10pt}
\setlength{\rightmargin}{0pt}
}
\item[$^{\rm a}$] Total flux density recovered in the image 
\item[$^{\rm b}$] Corresponds to the A component
\item[$^{\rm c}$] Corresponds to the B component
\end{list}
\end{footnotesize}
\end{table}

\begin{table}[htb]
\caption{Source characteristics on mas-scales}
\label{tab:characteristics}
\[
\centering
\resizebox{\columnwidth}{!}{%
\begin{tabular}{@{}c@{~\,}c@{~\,}c@{~\,}c@{~\,}cp{35mm}@{}}
\hline
\hline
\noalign{\smallskip}
 & & {\footnotesize Eastern} & {\footnotesize Western} & \\
 & {\footnotesize Mean} & {\footnotesize Jet} & {\footnotesize Jet} & {\footnotesize Gap} \\
  $\nu$        & {\footnotesize P.A.\ }            & {\footnotesize Length} & {\footnotesize Length} & {\footnotesize Width}            & {\footnotesize Notes} \\
 {\scriptsize [GHz]} & & {\scriptsize [mas]} & {\scriptsize [mas]} & {\scriptsize [mas]} &  \\
\noalign{\smallskip}
\hline
\noalign{\smallskip} 
5    & $65^\circ$ & 30     & 30      & $\sim$1.3               & A dominates; B totally absorbed  \\
8.4  & $65^\circ$ & 14     & 14      & $<$0.5         & A dominates; B occupies the 5\,GHz gap region \\
22   & $67^\circ$ & 13     & 6       & $\sim$0.5               & B dominates     \\
43   & $70^\circ$ & 5      & 3.5     & $\sim$0.2             & B dominates; western jet weak                   \\
\noalign{\smallskip}
\hline
\end{tabular}
}
\]
\end{table}

\subsection{Linearly polarized intensity imaging}
\label{pol}
\begin{figure*}[htb]
\vbox{\includegraphics[clip, width=12cm]{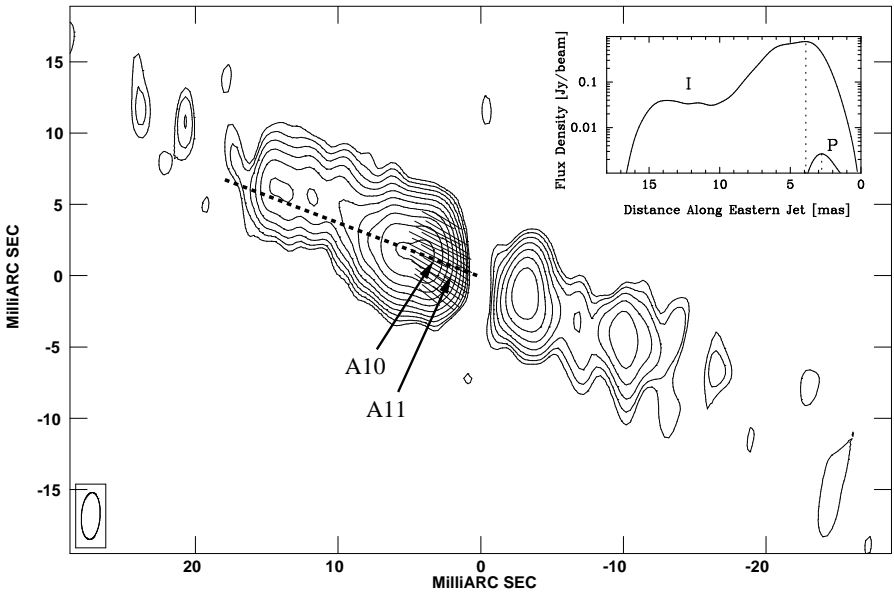}}\vspace{-2.5cm}
\hfill \parbox[b]{5.5cm}
{\caption{5\,GHz image of the polarized emission in NGC\,1052 { (EVPA 
superimposed on the total intensity)}. The inlayed panel
shows the total intensity and polarized emission { profiles along the
slice marked with the dashed line.}
              \label{fig:C_POL}}}
    \end{figure*}

Figure~\ref{fig:C_POL} shows the linearly polarized
intensity and its EVPA overlaid on the total intensity image at 5\,GHz.
A region of linearly polarized emission is visible at the base of the eastern
jet. The peak is
about 3\,mJy per beam and the EVPA is about 70$^\circ$, roughly parallel to the jet.
To decide whether the region of linearly polarized emission is resolved,
slices along the jet axis of both the total intensity image and
the polarization image were produced, using the task {\sc slice} in \aips\
(inlayed panel in Fig.~\ref{fig:C_POL}). In
this plot, the polarized emission peaks about 1\,mas offset
from the total-intensity maximum. A comparison to Fig.~\ref{fig:allmodels} ascribes
this part of the jet to component A\,10, which is optically thick at 5\,GHz.
The asymmetry of the polarized emission slice suggests the presence
of at least two components. The polarized emission is thus slightly resolved
and originates in the optically thick part of the eastern jet at 5\,GHz.

No linearly polarized emission at a flux density level above 1\,mJy 
could be detected at the other three frequencies 
Usually, the
degree of polarization rises with frequency since beam depolarization
reduces the degree of polarization at lower frequencies more strongly
than at higher frequencies.
Thus, one expects
the eastern jet to exhibit polarized emission from the region around
component A\,10 and A\,11 at 8.4\,GHz also. Assuming that the polarization has the same
(flat) spectrum as the total intensity in this region, it should reach 
$\sim 3$\,mJy per beam at 8.4\,GHz (comparable to the polarized
flux at 5\,GHz). However, this is not observed.
At higher frequencies
these components become optically thin 
(compare Sect.~\ref{spectral})
and thus fall in total intensity.
Were their emission polarized to the same degree as at 5\,GHz it would
be $< 1$\,mJy per beam and would thus lie below the detection threshold.
The non-detection of linear polarization at the high frequencies in our
observations is consistent with the results of Middelberg et al. 
(\cite{Mid04}) who report unpolarized emission from NGC\,1052 at 15\,GHz down to
a limit of 0.4\,\%.

\section{Identifying the center of activity}
\label{coreshift}
The symmetry between the jet and the counterjet constrains the position
of the central engine in NGC\,1052.
The cores of both jets 
are located at the distances to the central engine, $r_c$, where
the optical depth $\tau$ has fallen to $\sim1$. 
In a conical jet geometry
this distance is given
by:  $r_c \sim \nu^{-\frac{1}{k_r}}$, (see, e.g., Lobanov \cite{Lob98}) 
where $k_r=((2\alpha -3)b-2n-2)/(5-2\alpha)$, with 
$\alpha$ being the spectral index and
$b$ and $n$ the power
indices of the magnetic field and the density of the emitting particles:
$B \sim r^{b}$, $ N \sim r^{n}$ (Lobanov \cite{Lob98}).
Taking logarithms leads to:
\begin{equation}
\log(r_c) = -\frac{1}{k_r} \log(\nu) + const.
\end{equation}
Measuring $r_c$ at two frequencies allows one to determine
$k_r$ in the corresponding region of the jet.
For a freely expanding jet in equipartition
(Blandford \& K\"onigl \cite{Bla79}), $k_r$=1. The value of $k_r$ is larger in regions with steep pressure
gradients and may reach 2.5, for moderate values of $m$ and $n$ (Lobanov \cite{Lob98}).  
If external absorption determines the apparent core
position, comparable density gradients of the external medium can alter
$k_r$ to values above 2.5.

\begin{figure}[htb]
\centering
\includegraphics[width=\linewidth]{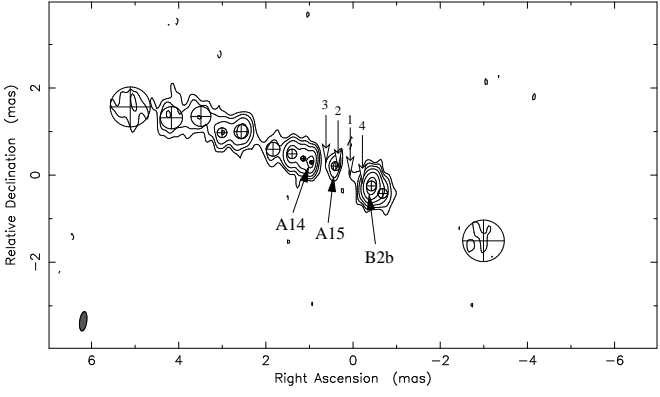}
\caption[Image with the model fitting results to the
43\,GHz data]{Image with the model fitting results of the
43\,GHz data. 
The innermost jet components are labeled.
The putative locations of
the central engine are indicated 
for four different scenarios (see discussion in the text).}
\label{fig:Qmodel}
\end{figure}

The values of $k_r$ deduced depend crucially on the absolute values of
$r_c$ on the two sides and therefore on the assumed position of the central
engine. Four scenarios have been tested with different reference points
(see Fig.~\ref{fig:Qmodel}). Table~\ref{tab:k_r} gives the derived values of
$k_r$ for each scenario. { In each case the position of component A\,14 
has been assumed as the core of the eastern jet, rather than A\,15. The latter
is comparably weak and, thus, most likely does not represent the true jet core.
A\,15 might rather represent a bright but heavily self-absorbed new jet component. 
Formally, however, the $k_r$ values for the scenarios 1,2, and 4 between 22\,GHz
and 43\,GHz change to $1.4\pm 0.06$, $0.6\pm 0.03$, and $1.0\pm 0.04$ if A\,15
instead of A\,14 is used. For scenario 3, A\,15 cannot be associated to the 
eastern jet core (compare Fig.~\ref{fig:Qmodel}).}

The area between the model components A\,15
and B\,2b is the most likely location of the central engine and the center
between both components is a natural choice for its exact position
(scenario 1). Shifting the
reference point eastwards (scenarios 2 and 3) alters the values of $k_r$
into unphysically large regimes (requiring density gradients $\propto
r^{-10}$ and higher). Assuming the true center of activity to be located
more westwards (closer to B\,2b, scenario 4), the values of $k_r$ derived
are still acceptable.

\begin{figure}[t!]
\centering
\includegraphics[width=\linewidth,clip]{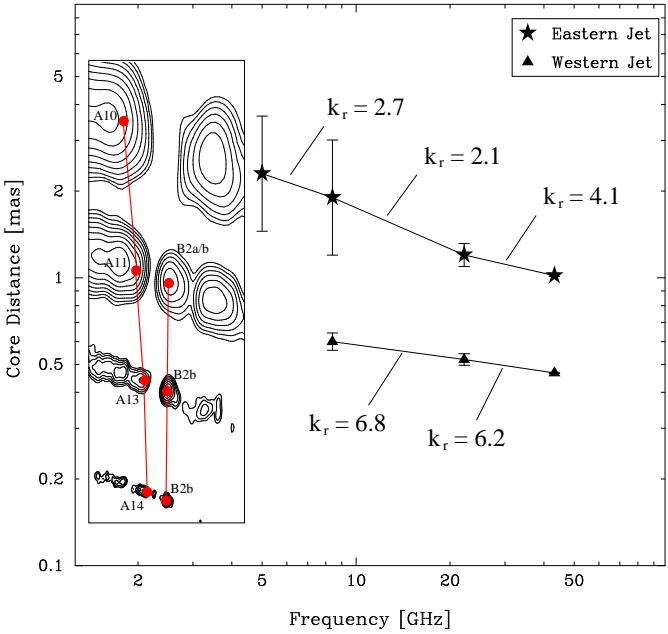}
\caption[Core positions in the two jets at the different
frequencies]{Core positions in the two jets at the different
frequencies for scenario 1 (see Fig.~\ref{fig:Qmodel}).
Table~\ref{tab:k_r} provides the values of $k_r$ for
the other { two} scenarios. The inset panel shows the core locations in
the source at the four frequencies. 
}
\label{fig:k_r}
\end{figure}

The positions of the bases of both jets at the different frequencies for
the first case (scenario 1) are shown in Fig.~\ref{fig:k_r}.  The eastern
jet has rather high values of $k_r$ below 22\,GHz, although still in
agreement with steep pressure gradients in the jet environment. Above
22\,GHz $k_r$ is 3.9$\pm$0.8, which is a good indicator for free-free
absorption affecting the jet opacity. The western jet has values of $k_r$
as high as 6.8$\pm$2.7 between 22\,GHz and 43\,GHz, suggesting a large
contribution from free--free absorption.

The results from the core shift analysis support the picture of a
free--free absorbing torus covering mainly the inner part of the western
jet and also a smaller fraction of the eastern jet. 
The true center of
activity in NGC\,1052 can be determined to lie between the model components A\,15 and B\,2b, with an uncertainty of only $\sim 0.03$\,pc.


\begin{table}[htb]
\caption{Values of $k_r$ for the four different putative centers of
activity}
\label{tab:k_r}
\[
\centering
\resizebox{\columnwidth}{!}{%
\begin{tabular}{@{}c@{\,}c@{~}cc@{~}cc@{~}c@{}}
\hline
\hline
\noalign{\smallskip}
& \multicolumn{2}{@{}c@{}}{(5--8.4)\,GHz}
& \multicolumn{2}{@{}c@{}}{(8.4--22)\,GHz}
& \multicolumn{2}{@{}c@{}}{(22--43)\,GHz} \\
\multicolumn{1}{@{}c@{}}{\scriptsize Scenario}
&~$k_{r,{\rm east}}$
&$k_{r,{\rm west}}$ 
&$k_{r,{\rm east}}$
&$k_{r,{\rm west}}$ 
& $k_{r,{\rm east}}$
& $k_{r,{\rm west}}$ \\ 
\noalign{\smallskip}
\hline
\noalign{\smallskip}
{\bf 1}
&2.7$\pm$1.9 &--     
&2.1$\pm$0.5 &6.8$\pm$3.0 
&4.1$\pm$0.8 &6.2$\pm$2.2 \\
{\bf 2}
&2.4$\pm$1.7 &--           
&1.8$\pm$0.4 &9.8$\pm$4.3 
&3.1$\pm$0.6 &9.3$\pm$3.4 \\
{\bf 3}
&2.1$\pm$1.5 &-- 
&1.4$\pm$0.3 &12.9$\pm$5.2 
&2.2$\pm$0.4 &12.5$\pm$4.6 \\ 
{\bf 4}
&3.0$\pm$2.1 &--          
&2.5$\pm$0.6 &3.7$\pm$1.6 
&4.9$\pm$1.0 &3.0$\pm$1.0 \\
\noalign{\smallskip}
\hline
\end{tabular}
}
\]
\end{table}

\section{Spectral analysis}
\label{spectral}
Spectral information can be derived from multi-frequency VLBI data in two ways.
The first approach is to use the knowledge
of the proper alignment of the total intensity images 
to derive images of the spectral index
between two adjacent frequencies. The second approach is to derive spectra of the
model fit components. The approaches are somewhat complementary. The latter
gives a handy number of component spectra, which can be analysed in detail,
whereas the spectral index imaging gives the full course of the spectral index
along the jet axis. Particularly, this yields spectral information at
parts of the jet that are not adequately represented by Gaussian 
model components. The results of
both approaches will be presented in this section.

\subsection{Spectral index imaging}
\label{spix}
\begin{figure}[t!]
\centering
\includegraphics[width=\linewidth]{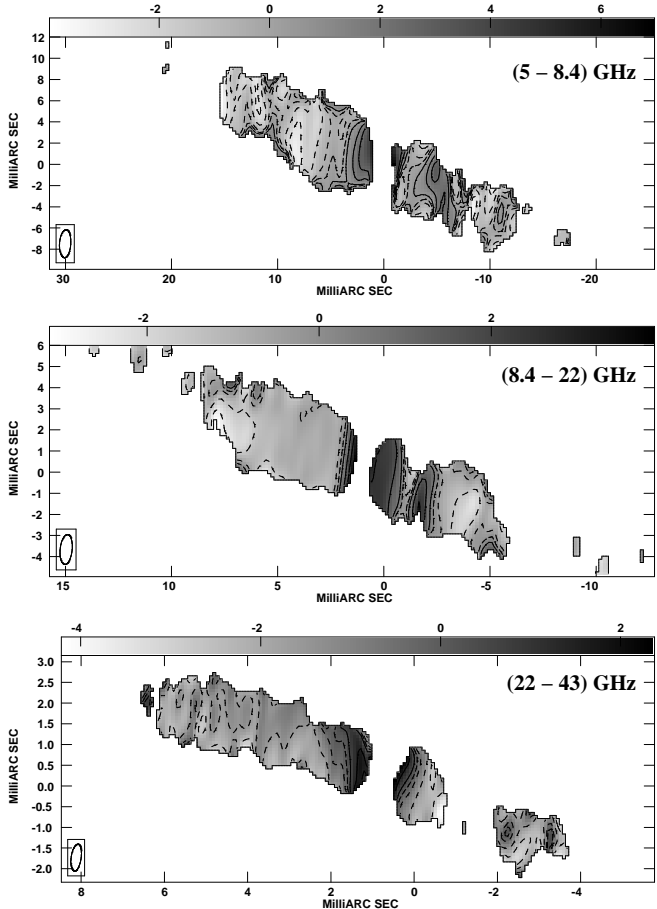}
\caption{Spectral index images of NGC\,1052. The contours
and beam parameters are given in Table \ref{tab:spixmaps}. Note the different scales
in the three images and that the images are centered on component B\,2. 
}
\label{fig:spix}
\end{figure}

Spectral index images have been produced using the \aips\ task 
{\sc comb}\footnote{At each pair of pixels of the two (coinciding) images,
the spectral index $\alpha$ ($S \propto \nu^\alpha$) 
is calculated from the flux densities per beam
$S_1$ and $S_2$ at the two frequencies $\nu_1$ and $\nu_2$: $\alpha = (\log{S_1} - \log{S_2}) 
/ (\log{\nu_1} - \log{\nu_2})$}.
For this, the total intensity has been reimaged with appropriate
tapering of the $(u,v)$-data to match the resolutions at adjacent
frequencies.
Information below 1\,mJy/beam was discarded in both input images. 
Table \ref{tab:spixmaps} gives the restoring beams and the other 
relevant parameters of the derived images. 
The spectral index images are shown in Fig.~\ref{fig:spix}. 

The main feature in each image of the spectral index is an optically thick inner edge of
the innermost part of both jets. Outwards along the jets, 
the spectral index tends to decrease.
The spectral index $\alpha$ exceeds the value of 2.5, the theoretical upper limit
for synchrotron self-absorption, on both sides of the gap in the (5--8.4)\,GHz
spectral index image 
and on the eastern side of the gap also between  8.4\,GHz and 22\,GHz. 
High values of
$\alpha$ are also reached on the western side of the gap in the (8.4--22)\,GHz
spectral index image 
and on both sides between 22\,GHz and 43\,GHz, which however do not exceed the critical
value of 2.5. Towards the outer jet regions the spectral index typically falls
to values $\lesssim -1$ in the whole frequency range between 5 and 43\,GHz. This
suggests that the emission from these outer jet regions far from the gap is
optically thin above 5\,GHz.

\begin{table}
\caption{Spectral index image parameters}
\label{tab:spixmaps}
\centering
\begin{tabular}{@{}cccc@{}}
\hline
\hline
\noalign{\smallskip}
$\nu_1$ &$\nu_2$ &Common Beam &Cutoff\\
{\scriptsize [GHz]} & 
{\scriptsize [GHz]} & 
{\scriptsize [mas$\times$mas,\,$^\circ$]} & 
\multicolumn{1}{@{}c@{}}{\scriptsize [mJy/beam]} \\ 
\noalign{\smallskip}
\hline
\noalign{\smallskip}
5 &8.4 &2.64$\times$1.06, $-$3.81 & 2  \\
8.4 &22 &1.42$\times$0.57, $-$5.75 & 1  \\
22 &43 &0.66$\times$0.24, $-$7.78 & 1  \\
\noalign{\smallskip}
\hline
\end{tabular}
\end{table}

\subsection{Spectral analysis of the model fit components}
\begin{figure}[t!]
\centering
\includegraphics[width=\linewidth]{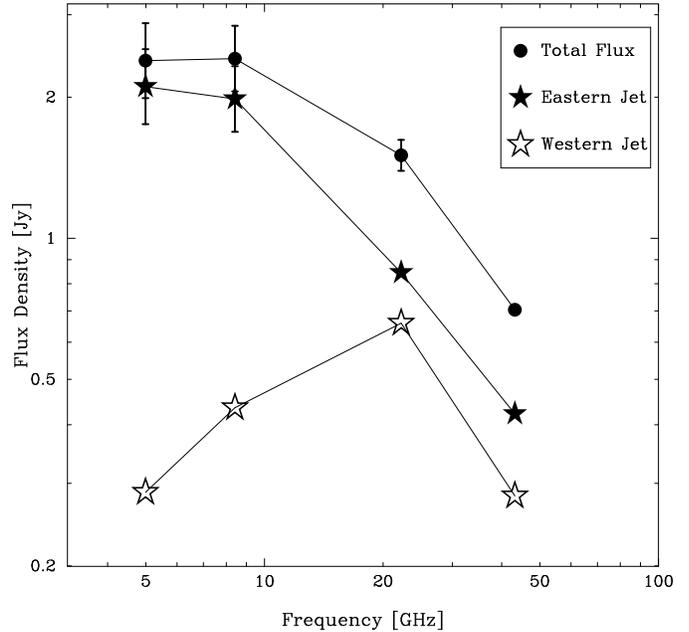}
\caption{Spectrum
of the parsec-scale structure of NGC\,1052 between
5\,GHz and 43\,GHz (total and for both jets separately). { Errorbars
are shown only if they exceed the symbol size.}}
\label{fig:spectrum1}
\end{figure}

The flux densities of the model fit components (see 
Figure~\ref{fig:allmodels}) 
can be used to derive spectra of {\sc i)}\,the whole parsec-scale
structure, {\sc ii)}\,the two jets separately and 
{\sc iii)}\,the model fit components
themselves. In Fig.~\ref{fig:spectrum1}, the total spectrum of the parsec-scale
structure of NGC\,1052 between 5\,GHz and 43\,GHz is shown, as well as the spectra
of both jets separately. All values were obtained by adding up flux
densities\footnote{We estimated the statistical standard uncertainty of all 
quantities derived from the model component
flux densities. The calculated
errors are shown in the plots, when they are not negligible.}
of model components (A\,1 to A\,15 for the eastern jet and B\,2b to D for the
western one).

At the high frequencies both jets show similar spectra with an optically thin
decrease of the flux density above 22\,GHz. The spectral index in this regime
is around $-1$ in both jets, with the eastern jet being significantly stronger
than the western one in agreement with the interpretation that the eastern jet 
approaches the observer
whereas the western jet is the counterjet. However, below 22\,GHz the
spectra of the two jets differs substantially. The eastern jet spectrum remains
optically thin above 8.4\,GHz and flattens around 5\,GHz.
The western jet,
on the other hand, exhibits a sharp decrease in flux below 22\,GHz.
Although the spectral
index limit of 2.5 is not exceeded, synchrotron self-absorption
 seems very unlikely to be responsible for the turnover of the
spectrum because of two reasons. First, the similarity of the spectra of
eastern and western jet at high frequencies suggests similar intrinsic
physical properties on both
sides so that the self absorption frequency should not differ by 
a factor of three. Second, the kinematical analysis of Vermeulen et al.
(\cite{Ver03}) shows the jet axis to be close to the plane of the sky and
the motions to be only weakly relativistic. There is little evidence for
strong Doppler boosting. Moreover, such an effect should shift the turnover
frequency of the counterjet to lower frequencies, rather than to higher ones.
In Sect.~\ref{orientation} we will use the moderate velocity differences,
on the one hand, and the
more pronounced brightness-temperature difference of jet and counterjet, on
the other hand, to
constrain the orientation of the jet-counterjet system. 

\begin{figure}[bt]
\centering
\includegraphics[width=\linewidth]{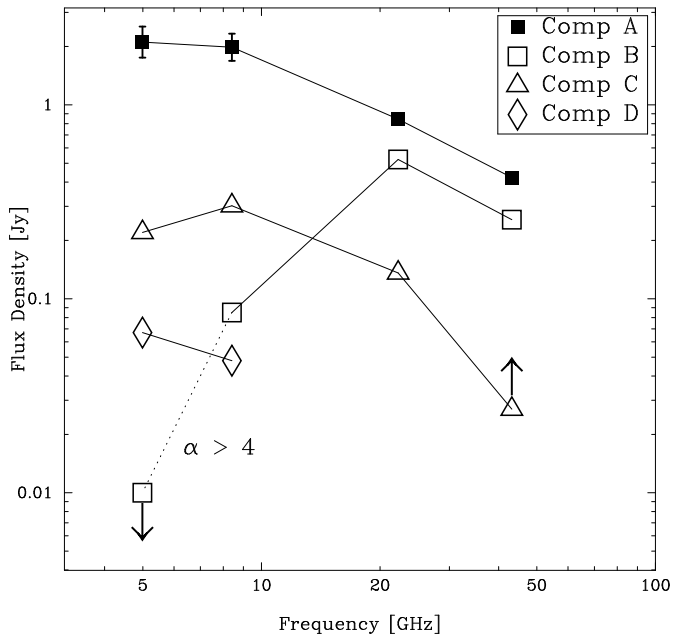}
\caption{Spectra of the
four jet regions A, B, C and D (see Fig.~\ref{fig:aligned} for the definition of
the jet regions). The arrows indicate that component C is partly resolved
at 43\,GHz { and component B is not detected at 5\,GHz. Thus, the shown values
represent} only a lower { and upper} limit{ , respectively}. 
{ Errorbars
are shown only if they exceed the symbol size.}}
\label{fig:spectrum2}
\end{figure}

If the spectrum of the western jet is decomposed into the spectra of the
different jet regions B, C and D (see Fig.~\ref{fig:aligned} and Fig.~\ref{fig:spectrum2}) it turns
out that the region B, the innermost region on the western side, is
responsible for the inversion of the western jet spectrum below 22\,GHz.
Its spectrum between 5\,GHz and 8.4\,GHz is highly inverted with
a spectral index $> 4$ since it is not detectable at all at
5\,GHz. Such 
a spectral index cannot be due to synchrotron self absorption
but indicates a region of external absorption towards the core of the
western, receding jet.
The highly inverted spectrum of this component in NGC\,1052 and the necessity of
an external absorber was first mentioned by Kellermann et al. (\cite{Kel99})
and later confirmed by Kameno et al. (\cite{Kam01}) and Vermeulen et al.
(\cite{Ver03}). 

It is not clear whether the eastern jet is also affected by
external absorption. Kameno et al. (\cite{Kam01}) 
proposed a geometry in which the obscuring torus, covers
0.7\,mas of the western jet and 0.1\,mas of the eastern jet.
Emission from A\,11 is not detected at 5\,GHz. Estimating (conservatively)
a flux density of 10\,mJy at 5\,GHz
yields a spectral index of $\sim 5$. This would suggest
that the innermost part of the eastern jet is also strongly affected
by external absorption. 
However, if { the flux densities of the two components A\,10 and A\,11} 
are summed at 5\,GHz and 8.4\,GHz, one
obtains a spectral
index of 2.3, which is close to that expected for
pure synchrotron self-absorption. { Consequently, due to the small separations 
between the inner
components in the eastern jet,} we cannot finally  judge
from this approach whether free-free
absorption plays an important role.
{ Kameno et al. \cite{Kam03} determined a spectral index of $3.28\pm 0.27$
for the base of the eastern jet between 1.6\,GHz and 4.8\,GHz and attribute this
to free-free absorption in the obscuring torus.} 

\begin{figure}[t]
\centering
\includegraphics[width=\linewidth]{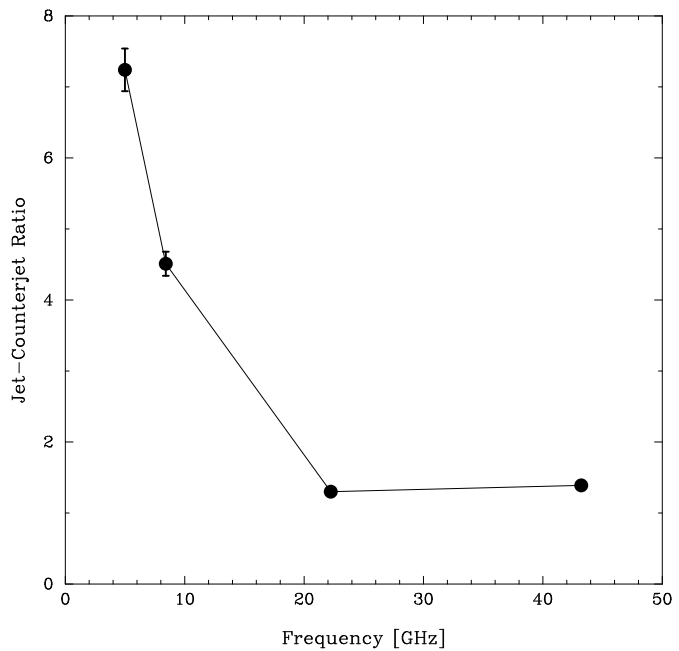}
\caption{The jet-to-counterjet ratio determined from the flux densities of the model fit components as a function of frequency. { Errorbars
are shown only if they exceed the symbol size.}}
\label{fig:jet_to_counterjet_ratio}
\end{figure}

The jet-to-counterjet ratio of NGC\,1052 can be determined
from the model fitted flux densities of both jets.
Figure~\ref{fig:jet_to_counterjet_ratio} shows the ratio of the flux densities of the
model components on either side of the gap as a function of frequency.
At high frequencies the jet-to-counterjet ratio is $\sim 1.5$ and starts rising
towards lower frequencies below 22\,GHz.
At 8.4\,GHz the jet is brighter than the counterjet by a factor of $\sim 5$ and 
at 5\,GHz the jet outshines the counterjet even by a factor of $\sim 7$. 
The jet-to-counterjet ratio is known to keep rising at lower frequencies up to
50 at 2.3\,GHz (Kameno et al. \cite{Kam01}). This suggests that a much bigger part
of the western jet is covered by the absorber 
than apparent from our high-frequency data.  
Alternatively,  curvature effects may play
an important role. 
In this case, the counterjet
bends away from the observer at a distance of a few milliarcseconds from the
core and its radiation is Doppler de--boosted.

\section{Brightness-temperature gradients}
\label{t_b}
In this section 
the brightness-temperature gradients along both jets in NGC\,1052 
are discussed. 
{ Starting from  the assumption that the}
magnetic field $B$, the electron density $N$, and the jet diameter $D$
can be described by power-laws:
\begin{equation}
B \propto r^{b}; N \propto r^{n}; D \propto r^d 
\label{gradients}
\end{equation}
it can be derived that
the brightness temperature, $T_{\rm b}$, is expected to fall 
with increasing distance from the jet base like:
\begin{equation}
T_{\rm b} \propto r^{s} \quad ,
\end{equation}
with $s<0$. { For this, one assumes optically thin synchrotron emission
with an emissivity 
$j_\nu \propto n_e B (\nu/\nu_B)^{\alpha}$
(Krolik \cite{Kro99}, Sect.\,9.2.1)}
and, implicitly, a constant Lorentz factor of the emitting electrons. 
Under these assumptions the brightness temperature distribution is determined
by the jet-geometry ($d=1$ for a conical jet, $d<0$ for a collimated jet, and
$d>0$ for a decelerating jet), the course of the magnetic field and
the particle density via
\begin{equation}
s= d + n + b(1-\alpha) \quad ,
\label{eq:s}
\end{equation}
where $\alpha$ is the spectral index. Typically, straight and continuous jets
exhibit values of $s \sim -2.5$ on parsec-scales (Kadler et al., in prep). 
The search for deviations from this power-law dependence provides a
tool to find regions in a jet which are affected by external
absorption or by abrupt changes of the jet parameters.

For a non-thermal source, the brightness temperature is a frequency-dependent
quantity (see e.g., Condon et al. \cite{Con82}):
\begin{equation}
T_{\rm b}=1.22\cdot10^{12} {\rm K} \left(\frac{S_\nu}{\rm Jy}\right)\left(\frac{\nu}{\rm GHz}\right)^{-2}\left(\frac{\Theta}{\rm mas}\right)^{-2} \quad ,
\label{eq:tb}
\end{equation}
with $S_\nu$ the flux density of the source, $\nu$ the observing frequency, 
and $\Theta$ the apparent diameter at half
maximum. In principle, the observing frequency has to be corrected for the
source redshift, but due to the small distance of NGC\,1052 ($z=0.0049$)
this correction is negligible.
According to
eq.~(\ref{eq:tb}), brightness temperatures have been computed for the model fit
components at all four frequencies. 
The uncertainties
in $T_{\rm b}$ were computed using Gaussian error propagation
from the errors in flux density $S$ and FWHM of the model components $\Theta$.
We assumed conservative values of 10\,\% for $\Delta S$ and 20\,\% for $\Delta \Theta$.

\subsection{The $T_{\rm b}$ distribution along the eastern jet}
\begin{figure}[t!]
\centering
\includegraphics[width=\linewidth]{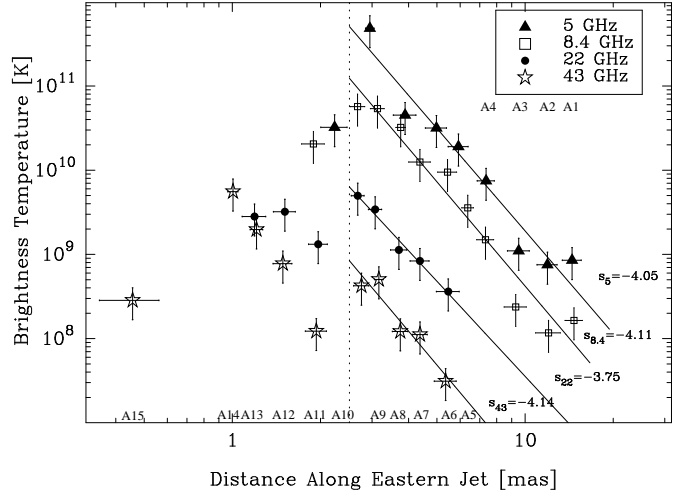}
\caption{Brightness temperature distribution along the eastern jet of NGC\,1052. Above the
cut-off distance of $\sim 2.5$\,mas the data can be fitted by a power-law (see Table
\ref{tab:tb_par} for the fit parameters) with a 
power-law index of
$\sim -4$. The offset between the four data sets of brightness temperatures reflects
the frequency dependence of $T_{\rm b}$ (see equation~\ref{eq:tb}).}
\label{fig:T_b1}
\end{figure}

Figure~\ref{fig:T_b1} shows 
the brightness temperatures in the eastern jet as a function
of distance from the central
engine. The latter was assumed to lie at the center between the components A\,15 and
B\,2b (see Sect.~\ref{coreshift} for a detailed discussion of the central engine
position in NGC\,1052).
$T_{\rm b}$ rises towards the
center following roughly a $r^{-4}$-law (compare Table~\ref{tab:tb_par}).
Between 2\,mas and 3\,mas from the central engine, however, 
there is an abrupt decrease of $T_{\rm b}$ (dashed line in Fig.~\ref{fig:T_b1}). 
This ``cut-off'' is present at all four frequencies although the
inner components (only visible at 22\,GHz and 43\,GHz) exhibit again a
rise in $T_{\rm b}$. The value of the
innermost component A\,15 falls
significantly below the extrapolation of the curve defined by A\,12, A\,13, and A\,14.

\begin{table}[b!]
\caption{
$T_{\rm b}$ gradient along the eastern jet
}
\label{tab:tb_par}
\begin{center}
\begin{tabular}{@{}ccc@{}}
\hline
\hline
\noalign{\smallskip}
$\nu$ & $\hat{T}_{\nu}$  & $s$  \\
{\scriptsize [GHz]} & {\scriptsize [K]} & \\
\noalign{\smallskip}
\hline
\noalign{\smallskip}
5   & $2.1^{+2.4}_{-1.1} \times 10^{13}$ & $-4.1 \pm 0.4$ \\
8.4 & $5.3^{+4.2}_{-2.3} \times 10^{12}$ & $-4.1 \pm 0.3$ \\
22  & $1.9^{+1.1}_{-0.7} \times 10^{11}$ & $-3.8 \pm 0.3$ \\
43  & $3.7^{+6.6}_{-2.3} \times 10^{10}$ & $-4.1 \pm 0.8$ \\
\noalign{\smallskip}
\hline
\end{tabular}
\end{center}
\begin{footnotesize}
Note:
We performed a power-law fit to $T_b$ beyond a distance from the
central engine $r=2.5$\,mas as
$T_{\rm b}(r) = \hat{T}_{\nu} \times r^s$, where
$\hat{T}_{\nu}$ is the extrapolated brightness temperature at 1\,mas and
$s$ is the power index.
\end{footnotesize}
\end{table}

\subsubsection{What causes the cut-off of the $T_{\rm b}$-distribution?}
We discuss three possible origins of the frequency--independent cut-off
of $T_{\rm b}$ in the eastern jet:
\paragraph{Free--free absorption:}
At a distance of $\sim 2.5$\,mas from the central engine 
the overhanging edge of the obscuring torus might
start to obscure a substantial fraction of the jet and thus reduce the brightness
temperature of its components via free-free absorption. 
If this interpretation was correct the pronounced frequency
dependence of the optical depth due to free-free absorption
($\tau_f \propto \nu^{-2.1}$) should be measurable: 
the free-free absorbed (i.e., observed) flux
density $S_{\nu ,{\rm abs}}$ depends on the intrinsic flux density $S_\nu$, the
optical depth of the absorber $\tau_f$ and the observing frequency $\nu$ as
\begin{equation}
S_{\nu ,{\rm abs}}=S_\nu \cdot e^{-\tau_f} {\rm \quad .}
\end{equation}
The calculated values of $\tau_f$ are given in Table \ref{tab:tau_f}
from which it is obvious that no pronounced effect of decreasing opacity with increasing
frequency is present. This makes the interpretation as free-free absorption
unlikely. 

\begin{figure}[t!]
\centering
\includegraphics[width=\linewidth]{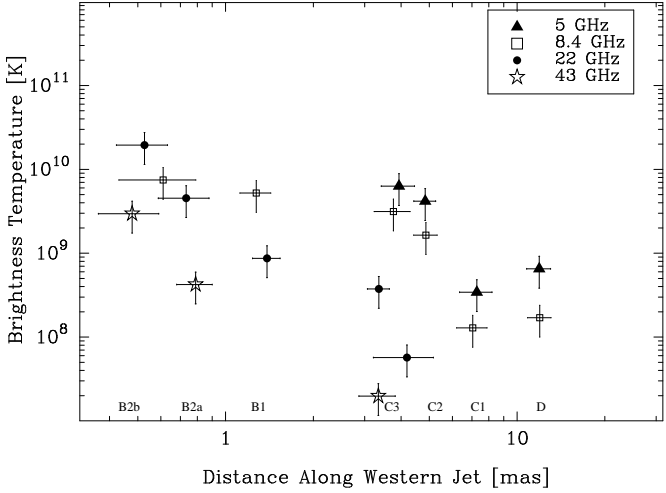}
\caption{Brightness temperature distribution along the western jet of NGC\,1052. No simple power-law can approximate the data, probably due to strong
curvature effects.
}
\label{fig:T_b2}
\end{figure}

\begin{table}[b!]
\caption{Optical depth values -- eastern jet
}
\label{tab:tau_f}
\begin{center}
\begin{tabular}{@{}lcccc@{}}
\hline
\hline
\noalign{\smallskip}
Id
& 5\,GHz   
& 8.4\,GHz 
& 22\,GHz  
& 43\,GHz  \\
\noalign{\smallskip}
\hline
\noalign{\smallskip}
A10    
& $3.2\pm1.1$
& ---
& --- 
& --- \\
A11 
& ---    
& $3.0\pm0.9$
& $2.5\pm1.0$ 
& $3.0\pm1.9$  \\
A12 
& --- 
& --- 
& $2.6\pm0.9$ 
& $2.3\pm2.3$ \\
A13 
& --- 
& --- 
& $3.6\pm1.1$ 
& $2.2\pm2.7$ \\
A14 
& --- 
& --- 
& --- 
& $1.9\pm3.3$ \\
A15 
& ---
& ---
& --- 
& $8.2\pm6.9$\\
\noalign{\smallskip}
\hline
\end{tabular}
\end{center}
\begin{footnotesize}
Note: Values for a distance from the central engine smaller than $r=2.5$\,mas
\end{footnotesize}
\end{table}

\paragraph{Synchrotron self-absorption:}
The synchrotron emission from a freely expanding, relativistic jet is
self-absorbed at distances smaller than 
\begin{equation}
r_{\nu} =\left[ \left(\frac{\nu}{K_{\rm
        jet}}\right)^{\epsilon +1} \frac{1}{\delta_{\rm j}^{\epsilon}
\phi_{\rm obs}} \right]^{1/(n+\epsilon\,b+1)}\,,
\label{eq:rnu}
\end{equation}
due to the change of the optical depth along the jet (Lobanov \cite{Lob98}):

\begin{equation}
\tau_s \propto r^{n+\epsilon b+1} \delta_{\rm j}^\epsilon\,. 
\end{equation}

Here, $\epsilon = 3/2 - \alpha$, $\nu$ is the observing frequency,
$\phi_{\rm obs}$ is the observed opening angle of the jet
(thus, implicitly assuming $d=1$; compare Equation~\ref{gradients}), $K_{\rm jet}$
is a bootstrap constant describing the jet conditions at a certain
characteristic distance (Lobanov 1998),  
and $\delta_j$ is the jet bulk Doppler factor.
Acceleration and { deceleration} of the
flow may affect the dependence of Equation~(\ref{eq:rnu}), and ultimately even cause the observed
abrupt decrease of the brightness temperature at shorter distances 
from the central engine (Fig.~\ref{fig:T_b1}). 

The effect of { deceleration} can
be accounted for by assuming $\delta_{\rm j} \propto r^{f}${ , with $f\le0$,} 
and $\phi_{\rm
  obs} \propto r^c$, with $c\ge0$. With these assumptions, the length of the
self-absorbed portion of the jet becomes $r_{\nu} \propto \nu^{\zeta}$, with
\begin{equation}
\zeta = \frac{\epsilon + 1}{\epsilon(b+f)+n+c+1} \,.
\label{eq:zeta}
\end{equation}

The peak-positions of the brightness temperature distribution in the
eastern jet (Fig.~\ref{fig:T_b1}) imply $\zeta \geq -1/10$ ({ note that} 
$\zeta < 0$) 
for the shallowest possible
slope of $r_{\nu}$. For a typical set of assumption about the jet flow ($b=-2$,
$n=-1$, $\alpha = -1$) this requires $f \leq -8 - 2/5 c$, which is implausible as
it implies extremely { strong deceleration} ($\delta_{\rm j} \propto r^{-8}$ and
higher) of the flow. This scenario can therefore be ruled out.

\paragraph{Steep pressure gradients:}
Frequency-independent local maxima of the brightness-temperature 
distribution along the jet can also be produced by
strong density and pressure gradients at the outer edge of the nuclear torus
and, to a lesser degree, by the gradients of the magnetic field strength
(Lobanov 1998). Such gradients may result in rapid changes of synchrotron
self-absorption and external free-free absorption of the jet emission, both of
them increasing the opacity at shorter distances from the nucleus.
These two factors together can in principle explain values of
$\zeta\ge -1/10$, if density gradients at the outer edge of the torus are
stronger than $n\approx -10$ ($N\propto r^n$). For the expected size of the
nuclear torus $r_{\rm tor} \sim 0.2$\,pc to $0.3$\,pc (corresponding to the distance
of 2\,mas to 3\,mas at which the brightness-temperature decrease sets on), and typical
densities in the absorbing torus ($n_{\rm H} \sim 10^6$\,cm$^{-3}$ to $10^{12}$\,cm$^{-3}$,
Cassidy \& Raine \cite{Cas93}) and in the nuclear medium of the host galaxy ($n_{\rm H}
\sim 10^2$\,cm$^{-3}$ to $10^{8}$\,cm$^{-3}$, Ferguson et al. \cite{Fer97}), the density increase
should occur on scales of $\sim 2.5 r_{\rm tor}$. This seems to
be the most plausible explanation for the observed brightness temperature
changes in the eastern jet.

\paragraph{Standing shocks:}
{ The approach to describe the jet parameters $B$, $N$, and $D$ with 
simple power laws (Equation~\ref{gradients}) assumes, implicitly, a 
quasi-stationary, continuous jet flow without perturbations. This model cannot
describe the influence of shocks in the jet flow on the brightness temperature
distribution. Qualitatively, the occurence of a cut-off in the 
$T_{\rm b}$-distribution at the location of a region of enhanced linearly polarised
emission (compare Sect.\ref{pol}) is in agreement with the expected observational
signature of a standing shock. Part of the bulk flow energy of the jet can be 
converted into magnetic energy of the jet plasma in such a standing-shock region.
This might cause $K_{\rm jet}$ to increase abruptly. Beyond the shock, the jet flow
might be quasi-stationary and thus well described by our model.}

\subsubsection{Derivation of the spectral index}
The frequency dependence of the brightness temperatures in the eastern jet
of NGC\,1052 (see Table \ref{tab:tb_par}) enables us to derive directly the
value of the spectral index $\alpha$: 
Equation~\ref{eq:tb} shows that the brightness temperature at a given
distance from the central engine depends on the spectral index as
$T_{\rm b} \propto \nu^{-2+\alpha}$, if optically thin synchrotron emission is
assumed. A linear regression of the brightness temperature at a distance of
1\,mas ($\hat{T}(\nu)$; see Table~\ref{tab:tb_par}) as a function of frequency
$\nu$ yielded $\hat{T}(\nu) \propto \nu^{-3.0 \pm 0.1}$. Thus, the 
spectral index is $\alpha = -1.0 \pm 0.1$. This is in good agreement
with the results of the spectral index imaging (see Sect.~\ref{spix}).

\subsubsection{The gradients of particle density and magnetic field}
The fitted sizes of the Gaussian components at the four frequencies do not
show significant deviations from a conical jet-structure so that $d=1$ can be
assumed. Applying this to Equation~(\ref{eq:s}) leads to
\begin{equation}
n + 2 b \sim -5 \quad .
\label{eq:n+b}
\end{equation}
In the most simple scenario of a conical expanding jet with a well defined
particle energy distribution function (i.e. without cooling; $n = -2$) 
and equipartition between magnetic energy and particle energy ($b = n/2$),
a relation $n+2b = -4$ is expected. Our result thus implies that at least
one of these two assumptions does not hold strictly. If one assumes that
equipartion holds, energy losses (e.g. due to adiabatic expansion) might steepen
the particle energy distribution effectively to $n \sim -2.5$.
An alternative explanation would be that the magnetic flux of a dominant
longitudinal component of the magnetic field density along the jet axis
($\pi D^2 B_z$) might be conserved so that $b = -2$. In the extreme case
of no transverse magnetic field component this would lead to $n + 2 b \sim -6$.

In principle, the detection of polarized emission in the eastern jet
(see Sect.~\ref{pol}) allows the direction of the dominant component
of the magnetic field to be determined. 
The linearly polarized emission originates in the
region around component A\,10, thus at the edge of the most compact component of
the obscuring torus. Faraday rotation in this region is expected to be large
so that we cannot directly derive a dominant transverse magnetic field from
the alignment of the EVPAs with the jet axis. Polarimetric observations
at multiple closely separated frequencies around 5\,GHz would be necessary to derive
the amount of Faraday rotation, which in combination with the measured 
opacity due to free-free absorption $\tau_{f}$ 
additionally would allow to disentangle
the particle density and the magnetic field (the Faraday rotation at a given
frequency is proportional to the particle density $n_e$, while $\tau_{f}
\propto n_e^2$).

\subsection{The $T_{\rm b}$ distribution along the western jet}
The brightness- temperature distribution along the western jet of NGC\,1052
is shown in Fig.~\ref{fig:T_b2}. Here, the situation
is more complex than on the eastern side. The western jet
is less continuous than the eastern jet, which might be due to stronger
curvature, particularly, between the regions C and D (see the counterjet 
structure at 22\,GHz in Fig.~\ref{fig:aligned}). 
Since the equations derived above assume a straight jet geometry without 
abrupt bends, we do not try to approximate the brightness-temperature distribution
by a single power-law. 
However, the course of the brightness temperature along the western jet
tells something about the orientation of the jets and
the structure of the obscuring torus towards the receding jet.

$T_{\rm b}$ decreases rapidly outwards between 3 and 12\,mas. At a distance of
about 7\,mas, however, in the area of component C\,1, 
the brightness temperature has a local minimum at 5\,GHz and 8.4\,GHz,
which might be due to Doppler de-boosting due
to a bend away from the line of sight. 
Inwards of 3\,mas, in region B, where the strongest effects of free-free
absorption were detected (see Sect.~\ref{spectral}) the $T_{\rm b}$ distribution is flatter
and the peak brightness temperatures do not exceed values of a few times
$10^{10}$\,K. 

Component B\,2 is strongly absorbed and, thus, 
its brightness temperature is substantially 
reduced at 8.4\,GHz, dropping even below the peak value at 22\,GHz (as expected
due to its inverted spectrum discussed above).
This can be used to estimate the absorbing column density towards component
B\,2, the area of the strongest effects of free-free absorption in NGC\,1052.
Assuming an intrinsic symmetry between jet and counterjet we can use the
measured frequency-dependence of
$\hat{T}(\nu)$ from the eastern jet of NGC\,1052 to estimate 
the intrinsic brightness ratio of the counterjet at 8.4\,GHz and 22\,GHz. 
Considering the slightly different fitted slopes (compare Table~\ref{tab:tb_par})
we derive from the best sampled region between 3\,mas and 8\,mas along the eastern jet that 
the brightness temperature at 8.4\,GHz should exceed
the value of $T_{\rm b}$ at 22\,GHz by a factor of $15$ to $20$.
Component B\,2 has approximately the same brightness temperature at 8.4\,GHz
and at 22\,GHz (mean value of B\,2a and B\,2b) which means that its flux density
at 8.4\,GHz is reduced by a factor of $\sim e^{-2.7}$--$e^{-3.0}$, i.e., 
$\tau_f^{8.4-22}({\rm B\,2}) \sim 2.7$ to $3.0$. 
{ This is in agreement with the results of Kameno et al. (\cite{Kam01}). 
These authors derive an optical depth of ~300 at 1\,GHz in the region corresponding
to our component B\,2, yielding $\tau_f \sim 3.4$ at 8.4\,GHz.}
The optical depth due to free-free absorption is given by
(e.g., Lobanov \cite{Lob98})
\begin{equation}
\tau_f = 30 \cdot 10^{16}\,L\,T^{-1.35}\,\nu^{-2.1}\,\overline{n}^2 \quad .
\end{equation}
For $\tau_f \sim 3$ at $\nu = 8.4$\,GHz at a temperature of $T=10^4$\,K,
and a length of the absorber of $L=0.3$\,pc
(comparable to the extent of the absorbing region in the plane of
the sky)
we derive a density of
$\overline{n}=2.4 \cdot 10^5$\,cm$^{-3}$ and an absorbing column density of
$2.2 \cdot 10^{22}$\,cm$^{-2}$ towards component B\,2.
Depending on the unknown ionization fraction of the torus material\footnote{{ 
Free-free absorption is an indicator of the column density $N_{\rm e}$ of 
free electrons in an ionised medium, rather than the column density of neutral
hydrogen $N_{\rm H}$. In a fully ionised medium, $N_{\rm e}=N_{\rm H}$ holds.}}
this value is consistent with various
X-ray observations of NGC\,1052, which imply a (model-dependent)
column density of $N_{\rm H}=10^{22}$\,cm$^{-2}$ to $N_{\rm H}=10^{24}$\,cm$^{-2}$ towards the unresolved nuclear
X-ray core (Weaver et al. \cite{Wea99};
Guainazzi et al. \cite{Gua99}; Kadler et al. \cite{Kad04}).
Because of the smaller absolute difference between the corresponding
values of $T_{\rm b}$ at 5\,GHz and 8.4\,GHz and the relatively larger uncertainties we
do not derive opacities for the outer components of the western jet. However,
from the inspection of Fig.~\ref{fig:T_b2} it is clear that the influence
of free-free absorption is weak beyond 4\,mas west of the nucleus. 

\subsection{Orientation of the jet--counterjet system}
\label{orientation}
The ratio of brightness temperatures in the jet and in the counterjet
at the same distances from the central engine can be used to constrain
the angle to the line of sight of the jet/counterjet axis:
\begin{eqnarray}
\frac{T_{b,j}}{T_{b,cj}}&=&\left(\frac{\delta_{j}}{\delta_{cj}}\right)^{2-\alpha} \quad , \\
\frac{\delta_{j}}{\delta_{cj}}&=&\frac{1+\beta \cos{\theta}}{1-\beta \cos{\theta}} \quad ,
\end{eqnarray}
where $\delta_{j}$ and $\delta_{cj}$ are the Doppler factor of the jet and
the counterjet, respectively.
The mean ratio of the measured values of $T_{\rm b}$ on the western side 
and the fitted value at the corresponding distance on the eastern side
(calculated for components
between 3\,mas and 5\,mas distance from the center where free-free absorption
effects are expected to be small) at all four frequencies
is $9\pm 2$. Assuming a spectral index of $-1$, this gives
$\beta \cos{\theta} = 0.35^{+0.03}_{-0.04}$. Since $\beta =1$ is an upper limit for
the jet speed this results in a maximum allowed angle of $\sim 72^\circ$.
The minimum allowed angle derived by Vermeulen et al. (\cite{Ver03}) 
is $\sim 57^\circ$ for which
$\beta = 0.64$. 
Thus, the jets are constrained to lie at an angle to the line of sight between
$57^\circ$ and $72^\circ$.

\section{Discussion and Summary}
\label{sum}
\begin{enumerate}
\item VLBI imaging of NGC\,1052 exhibits a parsec-scale ``twin-jet'' structure, matching 
the standard model of AGNs if the two jets are oriented
close to the plane of the sky.

\item We present accurately aligned, high-quality VLBI images of NGC\,1052 at
5\,GHz, 8.4\,GHz, 22\,GHz, and 43\,GHz, the associated spectral index images between the
adjacent frequencies and spectra of the various jet regions and model
fit components. 

\item The core of the western jet has a highly
inverted spectrum with a spectral index well above 2.5, the theoretical
upper limit for synchrotron self absorption, which was first mentioned by
Kellermann et al. (\cite{Kel99}) and later confirmed by Kameno et al.
(\cite{Kam01}) and Vermeulen et al. (\cite{Ver03}).
Qualitatively, we confirm the results of those
authors, particularly, the increasing opacity in the inner tens of
milliarcseconds of the eastern jet. 

\item We analysed the frequency dependence of the observed VLBI core position
in both jets and found another clear signature of free-free absorption
at the core of the western jet. The shift rate with frequency is too low
to be explained in terms of synchrotron self absorption alone, while the
core shift rate on the eastern side can still be explained under the
assumption of steep pressure gradients increasing the synchrotron opacity.
From the determination of these core shift rates we obtain an independent
measurement of the position of the central engine of NGC\,1052 with an
accuracy as high as $\sim 0.03$\,pc, superior to a kinematical derivation.

\item We find
a sharp cut-off of the brightness temperature distribution along the
eastern jet. Neither synchrotron self absorption nor free-free absorption
can explain this behaviour. The most plausible explanation is that we see
an effect of steep pressure gradients at a transition regime between the
external pressure-dominated jet regime and a more or less freely expanding
jet regime. Thus, the sharp cut-off of the brightness temperature distribution
marks an observational sign of the overhanging edge of the obscuring torus. 

\item We used the ratio of the observed brightness temperatures in the jet and
the counterjet to constrain the angle to the line of sight of the
``twin-jet'' system. Together with the information from the kinematical
study of Vermeulen et al. (\cite{Ver03}), the angle to the line of sight
can be determined to lie between $\sim$ 57$^\circ$ and $\sim$ 83$^\circ$.

\item The spectral index of the synchrotron jet emission was found to be 
$-1$. This result comes independently from the frequency dependence of the
brightness-temperature distribution in the eastern jet and from the imaged 
spectral index at large distances from the core.

\item Either equipartition between the magnetic energy and the particle energy
or the assumption of a single, well defined particle energy distribution
without cooling
is violated in the parsec-scale eastern jet of NGC\,1052.
Alternatively, a
conserved longitudinal component might dominate the magnetic field on these
scales.

\item We find a region of linearly polarized emission at the
base of the eastern jet. The EVPA of the polarized emission cannot be evaluated
directly since Faraday rotation at the center of this galaxy is expected
to be large. We find no linear polarization of the source at higher
frequencies. The simplest explanation for this behavior
is that the different
layers of the jet have a different degree of polarization and that the emission
at higher frequencies originates in an inner, unpolarized layer of the jet.
The higher column density at the more centrally located jet base at
higher frequencies could cause higher
Faraday depolarization than it does at 5\,GHz. This idea is
supported by the fact that the jet base at 5\,GHz coincides with the edge of
the obscuring torus.
This scenario can be tested with observations
at longer wavelengths and especially around 5\,GHz, where the depolarization
is expected to start dominating.

\item Most likely, a combination of free-free absorption, synchrotron self
absorption, and the presence of steep pressure gradients determine 
the parsec-scale radio properties of NGC\,1052.
An analytical model fit to the observed spectrum of only one single
absorption mechanism model seems not to be satisfactory to 
represent the true physical situation.

\item The absorbing column density derived from the degraded brightness
temperature of the western jet core is $\sim 2 \times 10^{22}$\,cm$^{-2}$,
in good agreement with the value obtained from X-ray spectroscopy
(Kadler et al. \cite{Kad04}).
This suggests that the nuclear X-ray emission (which is
unresolved for all X-ray observatories currently in orbit) originates on
the same scales which are imaged by VLBI and underlines the importance of
combined future radio and X-ray observations of NGC\,1052.

\end{enumerate}

\begin{acknowledgements}
We are grateful to T.\,Beckert, A.\,Kraus, T.\,P.\,Krichbaum, and
A.\,Roy for many helpful 
discussions and suggestions. We thank the referee,
Seiji Kameno, for his careful reading of the manuscript and his suggestions, which
improved the paper. M.\,K. was supported for this research through a stipend from the International
Max Planck Research School (IMPRS) for Radio and Infrared Astronomy at the University of
Bonn. 
\end{acknowledgements}

\end{document}